\documentclass{PoS}

\title{Background optimization for a new spherical gas detector for very light
WIMP detection}

\ShortTitle{Spherical Proportional Counter}

\author{\speaker{Ali Dastgheibi-Fard}$^a$, {I. Giomataris}$^{b}$, G. Gerbier$^{b}$, {J. Derré}$^{b}$, {M. Gros}$^{b}$, {P. Magnier}$^{b}$, {D. Jourde}$^{b}$, {E .Bougamont}$^{b}$, {X-F. Navick}$^{b}$, {T. Papaevangelou}$^{b}$, {J. Galan}$^{b}$, {G. Tsiledakis}$^{b}$, {F. Piquemal}$^{c}$, {M. Zampaolo}$^{c}$, {P. Loaiza}$^{c}$, {I. Savvidis}$^{d}$. {Saclay and LSM teams of - New Experiments With Sphere - network}\\
\llap{$^a$}LSM, Carr\'e  Sciences, 73500 Modane and CEA Saclay - IRFU/SEDI - 91191 Gif s Yvette\\
\llap{$^b$}CEA Saclay - IRFU/SEDI - 91191 Gif s Yvette\\
\llap{$^c$}LSM, Carr\'e  Sciences, 73500 Modane\\
\llap{$^d$}Aristotle University of Thessaloniki, Greece\\
\\E-mail: \email{ali.dastgheibi-fard@lsm.in2p3.fr}}

\abstract{The Spherical gaseous detector (or Spherical
Proportional Counter, SPC) is a novel type of particle detector, with a
broad range of applications. Its main features include a very low
energy threshold independent of the
volume (due to its very low capacitance), a good energy resolution, robustness and a single detection
readout channel, in its simplest version. Applications range from radon emanation gas
monitoring, neutron flux and gamma counting and spectroscopy to dark
matter searches, in particular low mass WIMP's and
coherent neutrino scattering measurement. Laboratories interested in
these various applications share expertise within the NEWS (New
Experiments With Sphere) network. SEDINE, a low background prototype
installed at underground site of Laboratoire Souterrain de Modane
is currently being
operated and aims at measuring events at very low energy threshold,
around 100 eV. We will present the energy calibration with
$^{37}$Ar, the surface background reduction, the measurement
of detector background at sub-keV energies, and show anticipated
sensitivities for light dark matter search.}

\FullConference{Technology and Instrumentation in Particle Physics 2014,\\
		2-6 June, 2014\\
		Amsterdam, the Netherlands}

\begin{document}

\section{Introduction}

There is an increasing interest for low-background, low-energy threshold detectors to identify the dark matter in our universe and to study  low-energy neutrino physics. 
The question of dark matter has indeed become essential to particle physics \cite{giomatarisNeutrino}.
The search for  WIMP  (Weakly Interacting Massive Particles) dark matter is under intense development and relies on the detection of low energy recoils (keV scale)  produced by the elastic interaction of WIMPS's with the nucleus of the detector. Increasing the sensitivity level requires detector scalable in target mass to 1 ton whilst maintaining the ability to reject backgrounds. The development of such  detectors remains a daunting challenge for nowadays and future low-background experiments. 
As for the light WIMP's (< 10 GeV), the nuclear recoil energy ($\sim$< keV) becomes extremely small, leading to a signal below threshold for most conventional solid or liquid state detectors. The challenge is to achieve a very low energy threshold, typically around  100 eV or below. 

Some new experimental ideas have come to maturity. Among them is the 
innovative  Spherical Proportional Counter (SPC), a gaseous detector, initially proposed by I. Giomataris \cite{giomataris2008novel}, which will allow to explore  a new region of dark matter particles of very low mass.  

Radioactive background studies about  the effect of the shield and radioactive contributions from used materials are necessary to understand and optimize the detection parameters. The low energy calibration and the reduction of the internal contamination of the inner surface of the detector and its effect on the detector background will be described below. 

\section{Detector description}

The detector consists of a large copper sphere (from 0.6 m to 1.3 m in diameter) and a small ball or sensor  (from 3 mm to 16 mm in diameter) located at the center of the vessel, 

\begin{figure}[htbp]
\begin{center}
    \begin{tabular}{cc}
      \includegraphics[width=12cm]{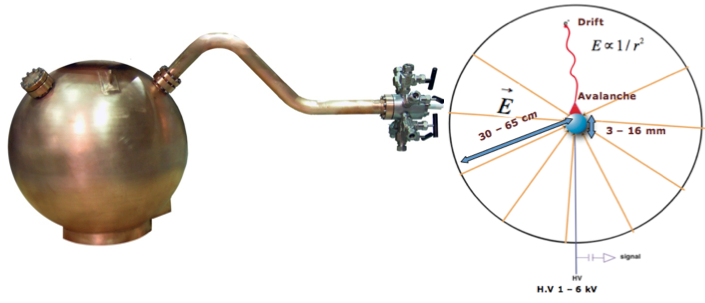}
         \end{tabular}
\caption{\textit{Left, 60 cm spherical detector and its tube for its operation from outside of shielding; right principle of operation of spherical gas detector }}
\label{Fig:DetectorAndElecField}
\end{center}
\end{figure}

thus forming a proportional counter. The ball is maintained at the center of the sphere by a metallic rod and is set at high voltage. The electric field varies in 1/r$^2$ and is highly inhomogeneous along the radius, allowing electrons to drift to the central sensor in low field regions constituting most of the volume, while they trigger an avalanche within few mm around the sensor. 

To homogenize the field inside the vessel, in some designs of the sensor, there is a second electrode along the rod at an optimized  distance  from the ball. The detector is operated in a sealed mode : the spherical vessel is first pumped out and then filled with an appropriate gas at a pressure of a few mbar up to 5 bar. The figure \ref {Fig:DetectorAndElecField} shows the detector and its operating principles. Detailed description of the detector, its operation and its performance can be found in the references \cite{Descrip1, Descrip2}. 

We can enumerate some advantages of the spherical detector such as : a) Low capacitance < 1 pF; b) Low energy threshold (< 1 keV); c) Good energy resolution; d) A single measurement channel for a large volume; e) Flexibility : gas, pressure and f) Robustness, simpleness and low price.


\section{Low energy threshold}
Our main concern for low energy calibration with a radioactive source was to perform a volume calibration with a short life time isotope.
$^{37}$Ar was selected as an isotope fulfilling the requirements. It is a gas, then filling the spherical volume and allowing to measure the homogeneity of response of the detector. It decays through electron capture with t$_{1/2}$ = 35 day, giving $^{37}$Cl which itself emits rearrangement single X rays X$_{k \alpha 1}$ of 2.6 keV and X$_{L \beta 3}$ of 260 eV.
It has the additional advantage that a weak source (few Hz) can be produced within our laboratory by irradiating $^{40}$Ca powder to a strong neutron source, by the reaction :
\[
^{40}Ca + n \rightarrow  ^{37}Ar + 3p
\]

The $^{37}$Ar source was  successfully tested for the first time with a spherical proportional counter at Saclay. The figure \ref{Fig:Calib37Ar} shows the calibration 

\begin{figure}[h]
 \centering
    \begin{tabular}{cc}
      \includegraphics[width=1 \linewidth]{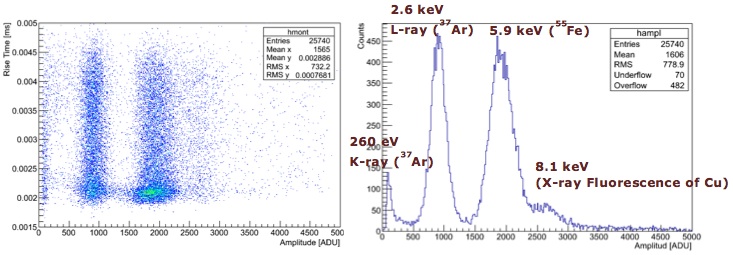}
      \\(a)~~~~~~~~~~~~~~~~~~~~~~~~~~~~~~~~~~~~~~~~~~~~~~~~~~~~~~~~~~~~~~~~(b)
     \end{tabular}
\caption{\textit{Low energy calibration at 50 mbar Ar + CH$_4$ (2\%) (a) rise time versus amplitude; (b) energy spectrum.}}
\label{Fig:Calib37Ar}
\end{figure}

of the spherical detector with $^{37}$Ar, together with $^{55}$Fe and X-ray fluorescence of the copper which is excited by ambient radioactivity.
The symmetric 2.6 keV peak (fig 2b) shows indeed a good homogeneity of response of the detector. The pulse rise time, determined by the longitudinal diffusion of the ionization electron along their path is related to the radius of the energy deposition. The rise time vs amplitude distribution (fig 2a) shows indeed the fiducialisation capability of the detector.

\section{Reduction of inner surface contamination}

SEDINE, a low activity 60 cm diameter prototype located at Laboratoire Souterrain de Modane (LSM), is dedicated to background studies in view of search for light dark matter at very low energy. In particular, due to the contamination of the inner surface of the detector by $^{210}$Pb and $^{210}$Po, the background  reduction  has been one of the main objectives since beginning of operation. Thanks to  three different chemical cleanings, we have been able to reduce the alpha rate from surface contamination by a  factor of $\approx$ 100 (fig 3a). 
For the first cleaning, we filled the half of the detector spherical volume   with a nitric acid solution at concentration above 17 \%, then turning 180$\,^{\circ}$, we cleaned the other half. Concerning the second cleaning, we sprayed the internal and external surface of the detector spherical volume with a nitric acid  solution at concentration above 30 \%. For the third cleaning, we used the drop by drop feeding method (of the same solution as the previous cleaning) because of the sensor sensitivity. After each cleaning with the nitric acid solutions, we used purified water to wash and hot nitrogen to dry the already cleaned parts

This allowed to obtain a background of a few events per keV.day  at a threshold of around 500 eV (fig 3b).

\begin{figure}[ht]
 \centering
    \begin{tabular}{cc}
   
      \includegraphics[width=1\linewidth]{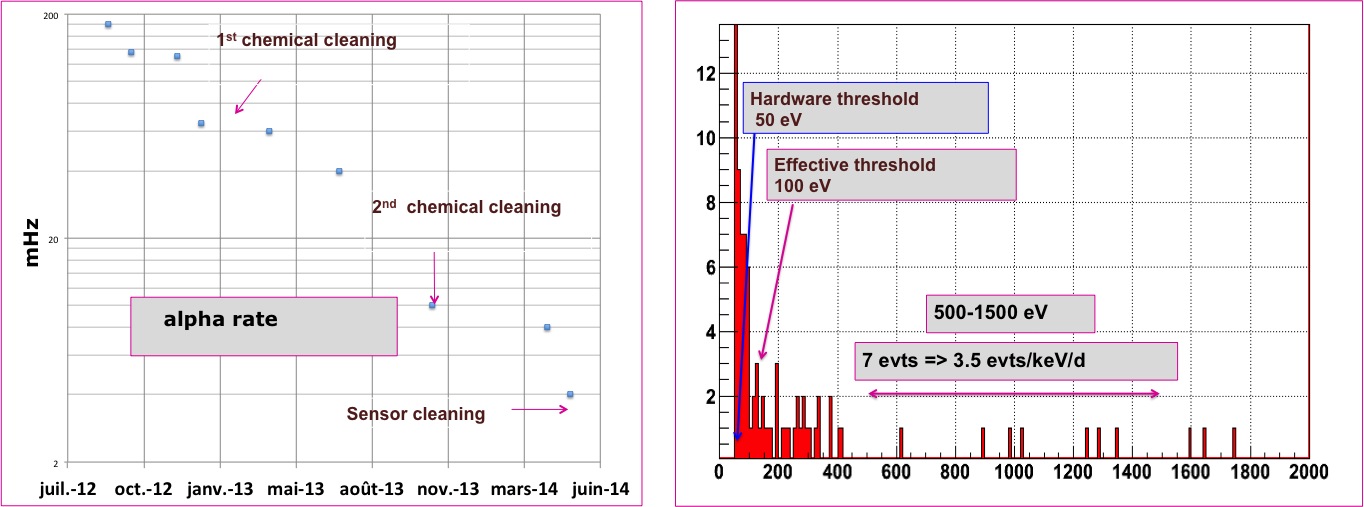}
      \\(a)~~~~~~~~~~~~~~~~~~~~~~~~~~~~~~~~~~~~~~~~~~~~~~~~~~~~~~~~~~~~~~~~~~~~~~~~~~~~~~~~~~~~~~~(b)\\
     \end{tabular}
\caption{\textit{(a) Reduction of the $\alpha$ rate after the different chemical cleanings (c) energy spectrum at very low energy with Ne+CH$_4$ (2\%).}}
\label{Fig:Calib37Ar}
\end{figure}

\section{Prospect}

One of the next goals within NEWS network  is to build a large detector operating with kg size target mass, with low mass nuclei and 100 eV threshold. The performed progress on background rejection and achieved level of radioactivity leads to anticipate competitive sensitivities for 
spin independent coupling WIMP's of masses below 8 GeV, using He and Ne nuclei as targets. More details on WIMP search and other applications can be found in  \cite{GGPaperNew}.
\\
\\
{\bf{Acknowledgements} }

We  wish to thank for their help the different NEWS network teams at Saclay, at Laboratoire Sourterrain de Modane  and at Laboratoire d'Annecy-le-Vieux de Physique des Particules. The low activity prototype operated in LSM has been partially funded by the European Commission astroparticle program ILIAS (Contract R113-CT-2004- 506222).


\begin{thebibliography}{99}
\bibitem{giomatarisNeutrino} I. Giomataris \textit{et al., Neutrino properties studied with a triton source and a large spherical TPC, Nucl. Instrum. Meth} A {\bf 530} (2004) 330-358.
\bibitem{SPCatLSM} I. Savvidis \textit{et al., Underground low flux neutron background measurements in LSM using a large volume (1m3) spherical proportional counter, Journal of Physics: Conf. Series.} Vol {\bf 203}. No 1. IOP Publishing, (2010).  
\bibitem{giomataris2008novel} I. Giomataris \textit{et al., A novel large-volume spherical detector with proportional amplification read-out,  JINST} {\bf 3} :P09007,(2008).
\bibitem{Descrip1} S. Aune \textit{et al, NOSTOS: a spherical TPC to detect low energy neutrinos, AIP Conf. Proc.} {\bf 785} 110-118 (2005).
\bibitem{Descrip2} I. Giomataris \textit{et al., NOSTOS experiment and new trends in rare event detection, Nucl. phys. Proc. Suppl.} {\bf150} 208-213 (2006).
\bibitem{GGPaperNew} G. Gerbier \textit{et al., NEWS : a new spherical gaz detector for very light WIMP detection} [arXiv:1401.7902 (2014)].




\end{thebibliography}
\end{document}